   \documentstyle[aps,epsf]{revtex}
    \catcode`\@=11
    \def\section{\@startsection{section}{1}{\z@}%
    {-3.5ex plus -1ex minus -.5ex}{1.5ex plus.3ex}{\bf }}
    \def\subsection{\@startsection{subsection}{1}{\z@}%
    {-3.5ex plus-1ex minus-.5ex}{1.5ex plus.3ex}{\bf }} %\makeatother
    
\begin{document}
\

    \vspace{2cm}

{\Large \bf
%%%%%%%%%%%%%%%%%%%%%%%%%%%%%%%%%%%%%%%%%%%%%%%%%%%%%%%%%%%%%%%%%%%%%%%%%% 
%%% here you should include the title of your contribution
Many-body spectral statistics of interacting Fermions
    }\vspace{.4cm}\newline{\bf   
%%%%%%%%%%%%%%%%%%%%%%%%%%%%%%%%%%%%%%%%%%%%%%%%%%%%%%%%%%%%%%%%%%%%%%%%%% 
%%% list of authors
M. Pascaud and G. Montambaux
    }\vspace{.4cm}\newline\small
%%%%%%%%%%%%%%%%%%%%%%%%%%%%%%%%%%%%%%%%%%%%%%%%%%%%%%%%%%%%%%%%%%%%%%%%%% 
%%% list of affiliations/institutions, please seperate each by a line 
%%% break via "\\"
Laboratoire de Physique des Solides, associ\'e au CNRS \\
Universit\'{e} Paris--Sud, 91405 Orsay, France
    \vspace{.2cm}\newline 
%%%%%%%%%%%%%%%%%%%%%%%%%%%%%%%%%%%%%%%%%%%%%%%%%%%%%%%%%%%%%%%%%%%%%%%%%% 
%%% the date follows here
%Received 1 April 1998, revised version 18 September 1998, accepted 6
%October 1998
\today
    \vspace{.4cm}\newline\begin{minipage}[h]{\textwidth}\baselineskip=10pt
    {\bf  Abstract.}
%%%%%%%%%%%%%%%%%%%%%%%%%%%%%%%%%%%%%%%%%%%%%%%%%%%%%%%%%%%%%%%%%%%%%%%%%% 
%%% the abstract follows here
We have studied the appearance of chaos in the many-body spectrum of interacting Fermions. The coupling 
of a single
state to the Fermi sea is considered. This state is coupled to a hierarchy of states corresponding to one or several 
particle-hole excitations. We have considered various couplings between two successive generations of this hierarchy and determined under which
conditions this coupling can lead to Wigner-Dyson correlations. We have
found that the cross-over from a Poisson to a Wigner distribution is characterized 
not only by the ratio $V/\Delta_c$, but also
by the ratio $\Delta_c/\delta$.  
$V$ is the typical interaction matrix element, $\delta$ is the energy distance between {\it many-body} states and
$\Delta_c$ is the distance between many-body 
states coupled by the interaction. 
    \end{minipage}\vspace{.4cm} \newline {\bf  Keywords:}
%%%%%%%%%%%%%%%%%%%%%%%%%%%%%%%%%%%%%%%%%%%%%%%%%%%%%%%%%%%%%%%%%%%%%%%%%% 
%%% please supply up to 4 keywords
%%% BEGIN CHANGE --------------------------------------------------------- 
Spectral correlations, Interactions 
%%% END CHANGE   --------------------------------------------------------- 
    \newline\vspace{.2cm} \normalsize
%%% BEGIN CHANGE --------------------------------------------------------- 
\section{Introduction}
Random Matrix Theory has been successful in describing the many-body
spectrum of interacting electronic systems\cite{Guhr98,Montambaux}.
An issue addressed recently is to understand how the transition from a Poisson
 to a Wigner-Dyson (WD) statistics occurs when interaction is switched on and what is the driving parameter for
this transition\cite{Berkovits,Shepelyansky97,Weinmann97,Jacquod97,Altshuler97}. The purpose of this work is to describe how the interaction drives the
appearance of chaos, i.e. how the  many-body  energy levels
can present WD correlations. We shall mainly
study the distribution $P(s)$ of nearest spacings $s$
between many-body levels. For WD correlations, it is well described by the Wigner surmise: $P(s)
= (\pi/2 s) e^{-\pi/4 s^{2}}$, where $s$ is the level spacing normalized to its
mean value.

As a frame of reference, we will consider 
 the recent work of 
Altshuler {\it et al}\cite{Altshuler97} which considers
the structure of the
many-body states when a single particle state interacts with a Fermi sea. 
Due to the interaction, this single particle state decays by the creation of an
electron-hole pair. The final state will be called a  $3$-particle state ($2$ electrons
and $1$ hole).
Because of the interaction, the initial state has a finite lifetime $\tau$.
In an infinite clean Fermi liquid, it is well known that the half-width of the
state $\Gamma=\hbar /2 \tau$ is proportional to $\epsilon^2/E_F$ where 
$\epsilon$ is the distance to the Fermi energy $E_F$.
In a diffusive system, the
effective interaction is increased since two diffusing quasiparticles have an enhanced
probability to interact several times\cite{Altshuler85}%
. As a result, the quasiparticle width is proportional to $\Delta (\epsilon
/E_{c})^{d/2}$ where $\Delta $ is the mean level spacing between quasiparticle
energies and  $E_{c}$ is the Thouless energy.

In a finite size system, an interesting new behavior arises when the
quasiparticle width becomes smaller than the level spacing $\Delta$. This
happens when $\epsilon <E_c$. In this case, individual particle peaks can be resolved%
\cite{Sivan}  and have been observed experimentally\cite{Sivan2}. In this
regime of non overlapping resonances, the width becomes dimension
independent and varies as $\Delta (\epsilon/E_c)^2$ \cite{Sivan}.
This description stays valid as long as the width is larger than the spacing
between final states $\Delta_3 =4 \Delta^3 /\epsilon^2$,  that is $\epsilon >
\epsilon^* \propto \sqrt{E_c \Delta}$, so that the Fermi
golden rule is applicable. In this regime, the final states, consisting of $3$-particle
states, are themselves unstable and can decay into a hierarchy of $%
5,\ldots, 2n+1$ quasiparticle states, also called states of generation $(2n+1)$. This  hierarchical structure of the Fock
space has been discussed by Altshuler {\it et al.}  who found interesting
properties of delocalization in this tree-like Fock space. One important issue concerns
 the ergodicity of the many-particle states and their spectral
statistics which is claimed not to be described by the WD 
approach\cite{Altshuler97}  (We will not consider here the regime $\epsilon
< \epsilon^*$ where the decay is not described by Fermi golden rule). This work has
also been reconsidered in more recent papers\cite{Silvestrov97,Mirlin97}.
\smallskip

Motivated by this  work, we have studied the statistical properties of
this many-body spectrum constituted of  hierarchical states coupled by the interaction matrix elements.
We assume that in the absence of
coupling, the levels obey  {\it Poisson} statistics. 
This would be the case for the single particle levels (1-particle states)
in a clean non chaotic cavity.  Even in the presence of disorder or in a ballistic chaotic
cavity, the 3-particle states are  described by a Poisson distribution: indeed, 3-particle states are formed by
addition of uncorrelated 1-particle states, since they have quite
different eigenenergies: 3-particle states and more generally $(2n+1)$-particle states thus 
 follow  Poisson statistics. The main goal of
this paper is to describe how level correlations evolve from Poisson to
WD as the interaction increases.

In order to describe how these correlations set in from the coupling between
states of the different generations, we have mainly studied the coupling
between two successive generations. First, we have considered the coupling
between one state (of generation 1) and a dense ensemble of energy levels
(generation 3). This is the well-known Bohr-Mottelson problem\cite{Bohr}. When the interaction is switched on, the 
 spacing distribution $P(s)$ deviates from a Poisson statistics and in the limit of
infinite coupling, it tends  to a limiting distribution which is intermediate between
Poisson and Wigner statistics. We study numerically how this distribution depends on the type of coupling. This is done in section II.
In section III, we consider the case where {\it several} states of generation 1 are  coupled to states of 
 the generation 3, considering that the
hierarchy stops at this generation, which is physically relevant when the energy resolution is limited at generation 3
by some inelastic broadening.  In this case,
 there is a cross-over
from Poisson to Wigner statistics which is driven by the number of intruders as well as the strength of the interaction. For a large number of intruders, the WD correlations appear when
the typical interaction matrix element becomes of the order of the spacing between final states. In section IV, we describe the spectral function (LDOS)
of an intruder state.

The case of generation 3 coupled to generation 5, or more generally the case of generation $(2n-1)$ coupled to
generation $(2n+1)$ (with $n \geq 2$) is more subtle. It is considered in section V. In this case, due to the two-body nature of the interaction,
the states of the generation $(2n-1)$ are only coupled to a {\it small} number of states of generation $(2n +1)$,
so that the number of final states   connected by the interaction is much smaller than the total number of
final states. It has been argued that in this case the cross-over to a WD statistics should
be uniquely dependent of the ratio $U/\Delta_c$ where $\Delta_c$
 is the distance between final states connected by the interaction\cite{Shepelyansky97,Weinmann97,Jacquod97}. We show that it is not true: the cross-over
 also depends on the density of final states.

\section{Coupling between  one level and a background}

We  first study the statistics of energy levels when one 
state (also called the intruder\cite{Flores98} or first generation state\cite{Altshuler97}) is coupled to  a dense ensemble of energy levels.
This ensemble will be also called the background or quasi-continuum, or
generation $3$ since in  ref.\cite{Altshuler97}, it consists of 3 quasiparticles
states. Their level spacing is  written $\Delta_3$. The typical coupling  strength is
written $U$. We will be interested in the regime validity of Fermi golden rule  where 
$U > \Delta_3$ or equivalently $\Gamma = \pi U^2 /\Delta_3 > \Delta_3$. Since we want to address this problem on
very general grounds, we will not refer more precisely to the problem studied in ref.
\cite{Altshuler97}. We assume that the states within a given generation
are not coupled directly (see comment in section V).
It is known  that when the background is described by a WD
statistics, a Gaussian coupling to an intruder does not change this
statistics\cite{Flores98,Bohigas}. The levels are shifted and the
correlation between old and new levels has recently been studied by Aleiner
and Matveev\cite{Aleiner}. Here, we want to see how the interaction
with the extra level can induce correlations between levels of the
background, starting with an original  {\it Poissonian} sequence.

Consider one
state $| \lambda \rangle$ , coupled to the background $\{|k\rangle\}$ of $N$
states obeying Poisson statistics via the Hamiltonian $H$: 
\begin{equation}
H_{0} = \epsilon_{\lambda} c_{\lambda}^{\dagger} c_{\lambda} + \sum_{k=1,N} \epsilon_{k}
c_{k}^{\dagger} c_{k} +
 \sum_{k=1,N} (V_{\lambda k} c_{\lambda}^{\dagger} c_{k} + h. c. )
\end{equation}
The "local" Green's function $G ^{R}_{\lambda \lambda} (E) = \langle \lambda ( |E - H + i0)^{-1} | \lambda \rangle$
 is  given by: 
\begin{equation}
G_{\lambda \lambda}^{R} (E) = (E - \epsilon_{\lambda}+i 0 - \sum_{k} \frac {%
V_{\lambda k} V_{k \lambda}} {E - \epsilon_{k}+i0}) ^{-1} \ \ .
\end{equation}
The imaginary part of this function defines  the Local Density of States (LDOS)
\begin{equation}
\rho_\lambda (E) =\sum_{|n\rangle} |\langle \lambda  | n \rangle|^2 \delta
(E-E_n)= -{\frac{1 }{\pi}} \mbox{Im} G_{\lambda \lambda}^{R} (E) 
\end{equation}
where $|n\rangle$ are the exact eigenstates, with energies $E_n$. This LDOS,
also called the strength function, describes the projection of the initial state 
$|\lambda\rangle$ on the eigenstates $| n \rangle$. When the background is indeed a continuum or when the energy
levels are broadened  by some mechanism, the imaginary part $\pi \sum_{k}
V_{\lambda k} V_{k \lambda} \delta(E - \epsilon_{k})$ can be replaced by:
$\Gamma = \pi U^2  / \Delta_3$
where $U = \sqrt{\langle V_{\lambda k}^2 \rangle }$ 
is the typical matrix element of the interaction and
$\nu_3=1/\Delta_3$ is the DOS of the background. Then the LDOS
has the Lorentzian shape: 
\begin{equation}
\rho_\lambda (E) = \frac{1}{\pi} \frac{\Gamma}{(E - \epsilon_{\lambda})^{2} +
\Gamma^{2}}\ \ .
\label{rholambda}
\end{equation}

However, when the states of the background are discrete, this Lorentzian is
just the envelope of a finer structure. This fine structure is obtained by
looking for the eigenstates $| n \rangle$ of the perturbed Hamiltonian,
which are the solutions of:

\begin{equation}
E - \epsilon_{\lambda} = \sum_{k} \frac {|V_{\lambda k}|^2} {E - \epsilon_{k}} \ \ .
\end{equation}

The right hand side of the equation, which is a function of $E$, diverges at
each $\epsilon_{k}$: in each interval $[\epsilon_k,\epsilon_{k+1}]$, there is exactly one
eigenstate $E_n$ (there are two more levels with energies $E <\epsilon_1$ and 
$E > \epsilon_N$ which tend to $\pm
\infty$ with $U$).

The levels near the intruder $\epsilon_\lambda$ are more perturbed than those far
from it. As a result the new statistics  depends on the position in the
spectrum. To avoid this
complication, we  first study the spectrum in the case of infinite
coupling $U \gg \Delta_3$ so that the spectrum is uniformly perturbed. The eigenstates are then the solutions of: 
\begin{equation}
\sum_{k} \frac {|V_{\lambda k}|^2} {E - \epsilon_{k}} = 0  
\label{sum0}
\end{equation}

We first consider the case of a {\it constant} coupling. The 
distribution $P(s)$ is shown on  fig. 1. It is intermediate between the
Poisson and Wigner statistics. The level repulsion occurs because each of
the new levels is locked between two levels of the original sequence $\{ \epsilon_k
\}$.
This distribution has been recently studied analytically by Bogomolny {\it %
et al.}\cite{Bogomolny98} who calculated the slope at small separation $s$: $
%P(s)= \pi {\frac{\sqrt{3}}{2}} s$.
P(s)= (\pi\sqrt{3}/2) s$.
\vspace{-0.5cm}

%------------------------------
\begin{minipage}[t]{6cm}
\begin{figure}
\centerline{
{\epsfysize=5cm\epsfbox{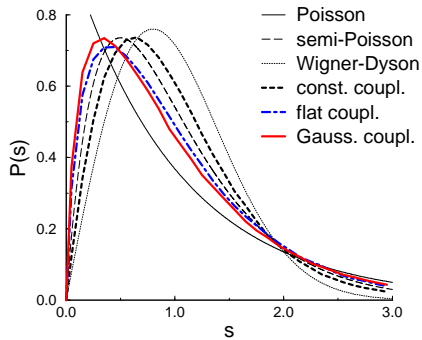}}}
\caption{\protect\small 
$P(s)$ for a Poisson background perturbed by a
constant, Gaussian or flat coupling to one intruder.
}
\label{fig2}
\end{figure}
\end{minipage}
\nolinebreak
\hspace{.5cm}
\begin{minipage}[t]{6cm}
\begin{figure} 
\centerline{
{\epsfysize=5cm\epsfbox{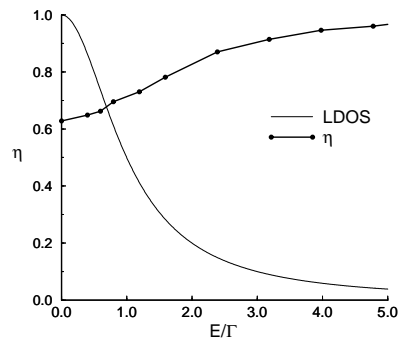}}}
\caption{\protect\small 
$\eta$ vs the distance $E/\Gamma$ to the intruder, in the case of Gaussian coupling. The thin line is the LDOS of
the intruder (in arb. u.) over the same energy interval. 
}
\label{etaspectre}
\end{figure}
\end{minipage}
%------------------------------
\medskip

To characterize this and other distributions throughout the paper,  we calculate the ratio \ 
$
\eta= (\int_0^{s_0} P(s) ds - \int_{0}^{s_0} P_{WD}(s)) / (\int_0^{s_0}
P_P(s) ds - \int_0^{s_0} P_{WD}(s))
$
where $s_0=0.4729$ is the first intersection point of the Wigner ($P_{WD}(s)$)
and Poisson ($P_{P}(s)$)
distributions\cite{Jacquod97}. 
$\eta$ interpolates between $1$
(Poisson) and $0$ (Wigner), see table I.
For the sake of comparing data, we also introduce another distribution obtained
if the sum (\ref{sum0}) would contain only pairs of neighboring levels $\epsilon_k$
and $\epsilon_{k+1}$.  In this case, the new eigenvalues would simply be given by $%
(\epsilon_k+\epsilon_{k+1})/2$. Since the $\{\epsilon_k\}$ follow a Poisson distribution, the new
distribution is:
$
P(s) = 4s e^{-2s}
$.
This distribution, called semi-Poisson, is also the distribution of nearest spacings
 for a plasma
model with short range logarithmic interaction\cite{Bogomolny98}.

\begin{center}
\begin{tabular}{|c|c|c|c|c|c|c|} 
\hline 
 WD& constant coupling&semi-Poisson&flat coupling&Gaussian coupling&Poisson \\ 
\hline
$0$ & $0.224$  & $0.386$  & $0.549$  & $0.639$ & $1$ \\ 
\hline
\end{tabular}
\end{center}

We have also studied the level distribution when the interaction $V_{\lambda
k}$ is not constant. We took two other probability distributions: i) a uniform (flat)
distribution over a finite interval.  
ii) a Gaussian
distribution.
We find that the repulsion is stronger with a constant coupling than for
flat or Gaussian couplings. A similar effect occurs in the case of an
intruder coupled to a GOE background\cite{Flores98}, in which case a
Gaussian coupling does not affect the level statistics significantly while a
constant coupling induces a quartic repulsion. 
The reason for this is the following: the eigenenergies  given by eq.(\ref{sum0}) are
trapped between the original levels $\{ \epsilon_k \}$. With a constant coupling, a
trapped level is repelled by two original levels with roughly the same
strength. On the contrary, with a random coupling, two eigenenergies are
closer to each other when the coupling is small and the repulsion is weaker.

When the coupling is finite, the spreading width $\Gamma$  is finite and the statistics of the eigenstates depends on the
 energy distance  $E$
to the intruder. At  $%
E \gg \Gamma$ the levels are weakly perturbed and their statistics remains close to Poissonian. On the opposite, when
$E$ stays smaller than $\Gamma$ the statistics is the   same as the one
obtained above when $U$ is infinite,  fig.(\ref{etaspectre}).

\smallskip

The nearest spacing distribution describes only short range level correlations.
To get some information on longer range correlations, we have also
calculated the number variance: $\Sigma^{2}(E) = \langle N^{2}(E) \rangle -
\langle N(E) \rangle^{2}$, which measures the fluctuation of the number of
levels $N(E)$ in a band of width $E$. Contrary to the WD case
 which is characterized by small fluctuations $\Sigma^{2}(E) \propto \mbox{ln} (E)$, 
here $\Sigma^{2}(E)$ shows large
fluctuations close to those of a Poisson distribution (fig. \ref
{plusintruderdeltan2}, solid thick line): for both constant and Gaussian couplings, $%
\Sigma^{2}(E)$ behaves like $E-1$ at large $E$. Since the new
states are locked between the initial states of the Poisson sequence, their
compressibility is the same as for the Poisson sequence, so 
that $\chi = \lim_{E \rightarrow \infty} \Sigma^{2}(E)/E \rightarrow 1$. 

To summarize this section, the coupling of a background with one extra level, 
 even when infinite, induces a cross-over from a Poisson to an
intermediate statistics which is still far from Wigner distribution. This is
because the   sequence of new levels
alternates with levels of a Poisson sequence.

\section{Coupling between several levels and a background}

We now study the case where several levels are coupled to the background.
Consider a set of $m$ intruders $\{\lambda_1,\lambda_2,\ldots,\lambda_m\}$ ($%
m \ll N$; $N$ is the number of states in the background) with  mean level spacing $\Delta$. As in the
case of one intruder, the Green's function can be calculated exactly. 
 It is the solution of the linear $m \times m$ system: 
\begin{equation}
[ (E I - H_{\lambda} - M ] \left( 
\begin{array}{c}
G_{\lambda_{1} \lambda_{1}} ^{R} \\ 
G_{\lambda_{1} \lambda_{2}} ^{R} \\ 
\vdots \\ 
G_{\lambda_{1} \lambda_{m}} ^{R}
\end{array}
\right) = \left( 
\begin{array}{c}
1 \\ 
0 \\ 
\vdots \\ 
0
\end{array}
\right)
 \ \ \mbox{with} \ \ 
M_{ij} = \sum_{k} \frac{V_{\lambda_i k} V_{k \lambda_{j}}} {E-\epsilon_{k} +i0}
\label{matrix}
\end{equation}
$H_{\lambda}$ is a diagonal matrix, whose diagonal elements are ${%
\epsilon_{\lambda_{1}}, \ldots, \epsilon_{\lambda_{m}}}$. $I$ is the identity matrix. The new  eigenstates are
given by the equation:
$det [(E I - H_{\lambda} - M ]=0$.
The problem has thus been reduced to the diagonalization of a $m \times m$
matrix.

Each intruder is broadened into a Lorentzian. It is clear that, as long as the width $\Gamma$ of each Lorentzian is small
compared to the distance $\Delta$ between intruders, the overlap between
them can be neglected and we are back to the previous problem of a single
intruder. This is the case as long as $\Gamma < \Delta$, i.e. when the
typical  interaction $U < \sqrt{\Delta \Delta_3}$  (In the case of the
disordered Fermi liquid where $U \simeq \Delta^2 / E_c$\cite{Sivan,Altshuler97,Blanter96} and $\Delta_3 \simeq
\Delta^3/\epsilon^2$, this means $\epsilon < E_c$\cite{Sivan}).
\vspace{-0.5cm}

%------------------------------
\begin{minipage}[t]{6cm}
\begin{figure} 
\centerline{
{\epsfysize=5cm\epsfbox{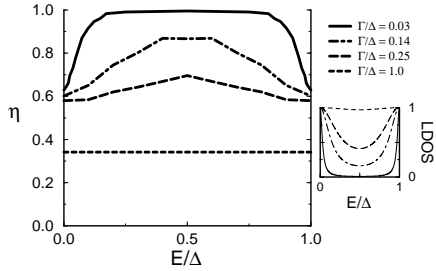}}}
\caption{\protect\small 
 $\eta$ vs $E/\Delta$, when $5$ intruders are coupled
to a Poisson background by a Gaussian coupling. The energy interval is
restricted between the  $2^{nd}$ and the $3^{rd}$ intruder.  Inset: Total spectral function 
$\rho(E)\equiv \sum_\lambda \rho_\lambda(E)$ for the same values of the coupling.}
\label{nooverlap}
\end{figure}
\end{minipage}
\nolinebreak
\hspace{.5cm}
\begin{minipage}[t]{6cm}
\begin{figure}
\centerline{
{\epsfysize=5cm\epsfbox{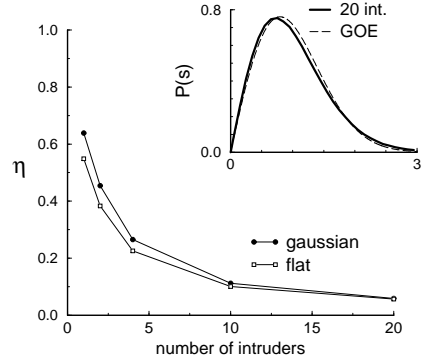}}}
\caption{\protect\small 
Variation of  $\eta$ with the number of intruders in the limit of infinite (Gaussian or uniform)  coupling. 
The inset shows
 $P(s)$ for the case of $20$ intruders, compared with the WD distribution.
}
\label{etamgene1}
\end{figure}
\end{minipage}
%------------------------------
\medskip

To characterize the spectral statistics, we  use the parameter $%
\eta$. A priori, it is a function of the position $E$ in the spectrum, of
the coupling strength $\Gamma$ and of the number $m$ of intruders: $%
\eta(E/\Delta,\Gamma/\Delta,m)$. When $\Gamma \ll \Delta$, the resonances do
not overlap and $\eta$ is a single function of $E/\Gamma$ as in the case of
a single intruder (Fig. \ref{nooverlap}). When the resonances overlap  enough
($\Gamma \gtrsim \Delta$), $\eta$ becomes almost energy 
independent: $%
\eta(\Gamma/\Delta,m)$ (Fig. \ref{nooverlap}). For infinite coupling, the statistics only depends on the number of
intruders: $\eta(m)$. As the number $m$ of these intruders increases, a
smooth transition towards the WD regime is now observed. Fig.(\ref{etamgene1})  shows
 how the parameter $\eta$ depends on $m$ and in the inset, it is shown that for $m=20$, $P(s)$ is
indeed very close to the  Wigner surmise. This smooth transition can also be
observed in the function $\Sigma^{2}(E)$. (fig. \ref{plusintruderdeltan2}).
\vspace{-0.5cm}

%------------------------------
\begin{minipage}[t]{6cm}
\begin{figure} 
\centerline{
{\epsfysize=5cm\epsfbox{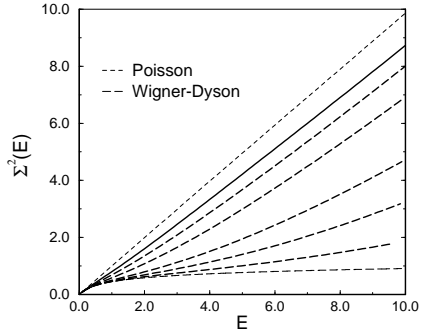}}}
\caption{\protect\small 
 $\Sigma^2(E)$ for a Poissonian background with an infinite coupling to $m= 1, 2, 4, 10, 20$ and $50$ intruders, for a Gaussian 
coupling, $\Gamma \gg \Delta$.
It approaches the WD  curve when $m$ increases. 
}
\label{plusintruderdeltan2}
\end{figure}
\end{minipage}
\nolinebreak
\hspace{.5cm}
\begin{minipage}[t]{6cm}
\begin{figure}
\centerline{
{\epsfysize=5cm\epsfbox{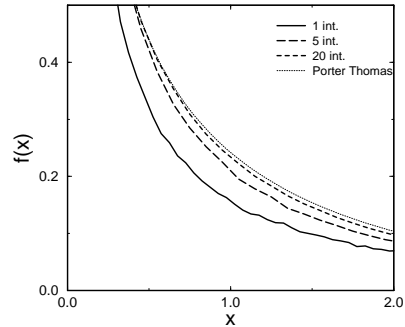}}}
\caption{
\protect\small 
Function $f(x)$ for a Poissonian background coupled to $m=1, 5$ or $20$ intruders. As $m$ increases, it
approaches the Porter-Thomas distribution 
%\cite{Porter}.
}
\label{strengthfn}
\end{figure}
\end{minipage}
%------------------------------
\medskip

We have also calculated the distribution for the coefficients of the
strength function: we compute the probability $f(x)$ for the weight $c_{n
\lambda} = |\langle \lambda | n \rangle|^{2}$ of an intruder $|\lambda \rangle
$ on an eigenstate $|n \rangle$ to be equal to $x$. The random matrix theory
predicts that $f(x)$, when properly renormalized should be equal to $f(x) =
1/\sqrt{2 \pi x} e^{-x/2}$ (for a time reversal symmetric system)\cite{Porter}.  We have  checked, that the function obtained is
the same for an infinite coupling, or for a finite coupling when the
eigenstates are taken in the central part of the Lorentzian.

Consider now the situation where the number of intruders is large  but the coupling is finite
$m \gg \Gamma/\Delta$. The statistics then  depends only on the overlap between
resonances: $\eta(\Gamma/\Delta)$ as it is shown on Fig.\ref{etaV10pf}.
Moreover the statistics of exact eigenstates do not depend on the relative
position of the intruders: we have checked that if the intruders have all
the same energy, or if they are randomly distributed, or if they are
regularly and equally spaced, the statistics of eigenstates is always the
same. 
\vspace{-0.5cm}

%------------------------------
\begin{minipage}[t]{6cm}
\begin{figure} 
\centerline{
{\epsfysize=5cm\epsfbox{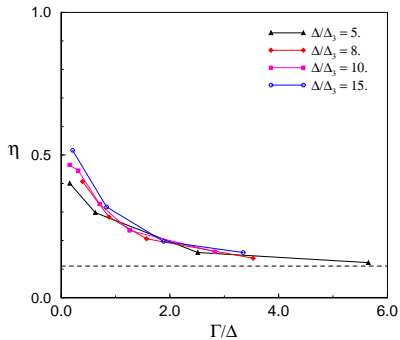}}}
\caption{\protect\small 
$\eta$ vs $\Gamma/\Delta$ for different distances $\Delta$ between 10 intruders equally spaced. 
The coupling is Gaussian. 
The horizontal dashed line shows the value of $\eta$ in the limit of infinite coupling for 10 intruders.
}
\label{etaV10pf}
\end{figure}
\end{minipage}
\nolinebreak
\hspace{.5cm}
\begin{minipage}[t]{6cm}
\begin{figure}
\centerline{
{\epsfysize=5cm\epsfbox{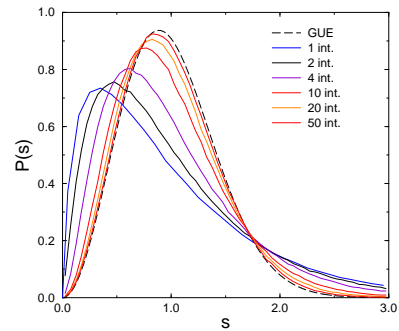}}}
\caption{\protect\small 
$P(s)$ for an infinite complex
coupling, as the number of intruders increases.
}
\label{gue}
\end{figure}
\end{minipage}
%------------------------------

\medskip

Finally, we have studied how the transition occurs when the coupling coefficients $V_{\lambda k}$ are complex, which corresponds to a  time reversal breaking situation. The 
 modulus of $V_{\lambda k}$ is chosen to follow a Gaussian distribution, and its phase follows a uniform distribution. As above we have considered the case of infinitely strong coupling.
As in the time reversal symmetric case, the transition depends on the number of intruders. 
Eq.(\ref{sum0}) shows that the case of one intruder 
is identical to the time reversal symmetric case
because the phases cancel.
In particular the nearest spacing distribution is linear for small spacings. 
On the contrary, when there are two or more intruders, the spacing distribution is quadratic for small spacings,
because the off-diagonal terms in eq. (\ref{matrix}) are now complex. 
As the number of intruders is increased, the distribution gradually evolves towards the  distribution  given by a random Gaussian {\em unitary} ensemble.

To conclude, we stress that the cross-over to WD statistics is induced by indirect terms $M_{ij}$, with $i \neq j$,
in the
matrix (\ref{matrix}). These off-diagonal terms  of the form 
$V_{\lambda_i k}V_{k \lambda_j}$  
are missing in a Cayley tree representation of
the hierarchical structure considered in ref.\cite{Altshuler97}. They are however essential for a correct description of the
spectral correlations.

\section{Spectral function in the case of several intruders}

Up to now, we have only considered the statistics of the eigenvalues. 
We now consider the LDOS of  the intruders. The LDOS for a given intruder $\lambda_1$, is given by the imaginary 
part of $G_{\lambda_1 \lambda_1}$. The matrix elements $M_{ij}$ in eq.(\ref{matrix}) are replaced by
their imaginary part (neglecting a shift of order $\Delta_3$ due to the real part):
\begin{equation}
M_{ij} =i \Gamma_{ij} =  i \pi \sum_{k} V_{\lambda_{i}k} V_{k \lambda_{j}}
\delta(E-\epsilon_{k})
\end{equation}
which, as long as the density of the background levels is much larger that the density of intruders,
 can be approximated by:
$
M_{ij} =-i \Gamma_{ij} = -i \pi \langle V_{\lambda_{i}k} V_{k \lambda_{j}} \rangle \nu_3
$
The eq.(\ref{matrix})  reads:
\begin{equation}
 \left( 
\begin{array}{ccc}
E-\epsilon_{\lambda_1} + i \Gamma_{11} &  & i \Gamma_{ij} \\ 
& \ddots & \\ 
i \Gamma_{ij} &  & E-\epsilon_{\lambda_m} + i \Gamma_{mm}
\end{array}
\right) \left( 
\begin{array}{c}
G_{\lambda_{1} \lambda_{1}} ^{R} \\ 
G_{\lambda_{1} \lambda_{2}} ^{R} \\ 
\vdots \\ 
G_{\lambda_{1} \lambda_{m}} ^{R}
\end{array}
\right) = \left( 
\begin{array}{c}
1 \\ 
0 \\ 
\vdots \\ 
0
\end{array}
\right)
\label{MMij}
\end{equation}
whose solution is:
\begin{equation}
G^R_{\lambda_{1} \lambda_{1}} =
\sum_\alpha  
{ 
\langle \lambda_1 | \alpha \rangle  \langle \alpha | \lambda_1 \rangle 
\over
E - E_\alpha
}
\end{equation}
The complex eigenvalues $E_\alpha$  give the position and the width of the resonances. We now consider several cases:

1) When the  coupling to intruders is uncorrelated and symmetric so 
that 
$\langle V_{\lambda_i k}V_{k \lambda_j} \rangle =
\langle V_{\lambda_i k} \rangle \langle V_{k \lambda_j} \rangle=0 $ 
for $i \neq j$, 
the off-diagonal elements in eq.(\ref{MMij})
vanish and one finds that
 each level coupled to the continuum
is broadened into a Lorentzian according to the Fermi golden rule; neither its center nor
its width are altered:
``the resonances do not talk to each other''.

2) When the coupling is such that 
$\langle V_{\lambda_i k}V_{k \lambda_j} \rangle =
\langle V_{\lambda_i k} \rangle \langle V_{k \lambda_j} \rangle
\neq 0 $, 
the off-diagonal elements of the matrix $M$ are now finite and  the resonances are  coupled.
 Given the form of the off-diagonal matrix elements, the eigenvalues can be calculated easily and found to be solutions of:
\begin{equation}
\sum _j{
 \gamma_j \over
E_\alpha - E_{\lambda_j}
+i(\Gamma_{jj}-\gamma_j)
}
= i
\label{solutions}
\end{equation}
$\Gamma_{jj}=\pi \nu_3 \langle V_{\lambda_i k}V_{k \lambda_j} \rangle$ 
are the width of the uncoupled resonances and
$\gamma_j=\pi \nu_3 
\langle V_{\lambda_i k} \rangle \langle V_{k \lambda_j} \rangle$.

3) Consider first the very specific case where $\Gamma_{jj}=\gamma_j$. This case corresponds to taking a constant coupling $V_{\lambda_j k}=V_j$  (which may still depend on the intruder $j$). It has been studied recently by K\"onig {\it et
al.} for the case of two intruders, e.g. two discrete levels of a quantum dot coupled to an electron reservoir\cite{Konig98}. The eigenvalues are solutions of
eq. (\ref{solutions}) with $\Gamma_{jj}=\gamma_j$
When the $\gamma_j$ increase and become large compared to the distance between intruders $\Delta$,
the resonances are split into several peaks whose centers are given by:
$
\sum _j
(\Re{E_\alpha} - E_{\lambda_j} )^{-1}
= 
0
$
and are placed at intermediate positions between the unperturbed levels.
Thus the LDOS $\rho_{\lambda_1}$ consists in a series of $(m-1)$ peaks centered on the positions $\Re{E_\alpha}$ and whose widths are given by $( \sum_j \gamma_j/(E_{\lambda_j}-E_\alpha)^2)^{-1} \simeq \Delta^2 / \gamma$.
The corresponding spectral weights are:
\begin{equation}
|\langle \lambda_1|\alpha \rangle|^2=
{1 \over (E_\alpha-\epsilon_{\lambda_1})^2} 
{1 \over \sum_j {1 \over (\epsilon_{\lambda_j} - E_\alpha)^2}} 
\end{equation}

The $(m-1)$ peaks actually carry  $(1-1/m)$ of the spectral weight.
The  rest is carried by a very broad Lorentzian of width $\sum_j \gamma_j$. 
This generalizes the case $m=2$ studied in ref.\cite{Konig98} where , at large coupling, a peak of spectral weight
$1/2$ appears in the middle of the two original discrete levels.
More generally, fig.(\ref{spectralweight}.a) shows that the spectral weight tends to concentrate at positions
placed between the original states.
\vspace{-0.5cm}

%------------------------------
\begin{minipage}[t]{6cm}
\begin{figure}[h]
\centerline{
\epsfysize 5cm
\epsffile{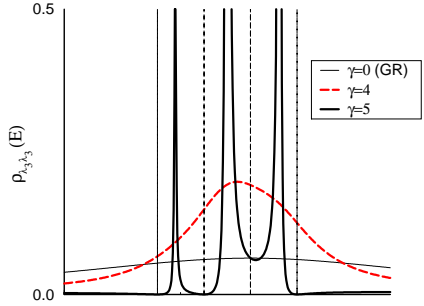}}
\caption{\protect\small 
LDOS of the $3^{rd}$ out of 4 intruders whose unperturbed energies are figured by strait lines. All $\Gamma_{jj}$'s and $\gamma_j$'s are equal. 
$\Gamma_j=5$, while $\gamma_j=0$ (Golden Rule case eq.\ref{rholambda}), 4 or 5.
}
\label{spectralweight}
\end{figure}
\end{minipage}
\nolinebreak
\hspace{.5cm}
\begin{minipage}[t]{6cm}
\begin{figure}[h]
\centerline{
\epsfysize 5cm
\epsffile{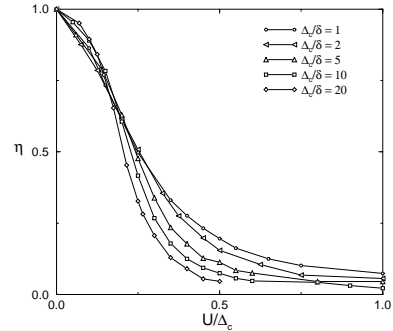}}
\caption{\protect\small 
$\eta$ vs $U/\Delta_c$ when the Fock space is restricted to the the last 2 generations $(2n^*-1)$ and $(2n^*+1)$. $\eta$ is also a function of the DOS 
$\delta$.
}
\label{PS_echelle}
\end{figure}
\end{minipage}
%------------------------------
\medskip

4) In the more general case where $\gamma_j \neq \Gamma_{jj}$,
the spectral function is still the superposition of $(m-1)$ peaks, centered at the same positions, but whose widths are 
$\mbox{Max}((\Gamma_{jj} - \gamma_j), \Delta^2/\gamma_j)$.
If the fluctuations of the coupling distributions of the couplings 
$V_{\lambda_j k}$ are sufficiently large so that 
$\Gamma_{jj}-\gamma_j \gg \Delta$, the peaks will transform into one single Lorentzian, whose width is however smaller than 
the width $\Gamma_{jj}$ predicted by the Golden Rule (eq.\ref{rholambda}).

5) In the most general case,  
$\langle V_{\lambda_i k}V_{k \lambda_j} \rangle \neq
\langle V_{\lambda_i k} \rangle \langle V_{k \lambda_j} \rangle$, 
the matrix $M$ has to
be diagonalized numerically.
In view of the preceding arguments we still expect the width of the resonances to be smaller than the bare resonance width.

\section{Coupling between generations of higher order}
We now consider the general case of a coupling between two generations $(2n-1)$ and $(2n+1)$ in the hierarchy of states
considered in ref.\cite{Altshuler97}. The spacing between
levels
in a given generation is\cite{Silvestrov97,Mirlin97}:
$
\Delta_{2n+1}=(2 n)!n!(n+1)!\Delta^{2n+1}/ \epsilon^{2 n}$. However, due to the two-body nature of the interaction, a level of generation $(2n-1)$ is coupled to a small number of 
states of the next generation. The distance between levels of generation $(2n+1)$ connected by the interaction is of
order:
$\Delta_c \simeq n \Delta_3 \simeq 4n \Delta^3/\epsilon^2$.
The direct coupling between states of a given generation can be neglected since the distance between such states
 is $\Delta_i \simeq \Delta^2/\epsilon$ which is always much larger than $\Delta_3$.
We now determine the condition under which the many-body states constituted of these two generations obey  WD statistics.
If this condition is obeyed the many-body states obtained by successive couplings from
 generations $(1)$ to $(2n +1)$ will also 
obey a WD statistics.
As noticed in ref.\cite{Altshuler97}, the DOS in a generation $(2n +1)$ is much larger than in the previous
generation. Due to the finite size of the system, this is true up to $n^* = \sqrt{\epsilon/ 2 \Delta}$. 
Then further generations are coupled to generations with smaller density of states, which certainly do not affect the appearance
of WD correlations.
We are thus especially interested in the two generations $(2n^*-1)$ and $(2n^*+1)$ which have the same level spacing $\delta \simeq
\Delta e^{- 4 n^*}$. Typically the distance between connected states is $\Delta_c \simeq \Delta/n^{*3} \simeq \Delta
(\Delta / \epsilon)^{3/2} \gg \delta$.
	
A priori three parameters are relevant, the inter-level spacing $\delta$, the spacing between states connected 
by the interaction $\Delta_c$
and the typical matrix element of the interaction $U$. It has been 
argued that the mixing of the states and the 
cross-over to a WD statistics depend only on the ratio $U/\Delta_c$
\cite{Shepelyansky97,Weinmann97,Jacquod97,Mirlin97}. 
Then, the cross-over is expected to occur when this
ratio is of order unity. Since $U \simeq \Delta /g$ where $g$ is the dimensionless conductance $E_c/\Delta$, this gives $\Delta/g \simeq \Delta (\Delta / \epsilon)^{3/2}$ so that $\epsilon \simeq g^{2/3}$ 
as found originally by Jacquod and Shepelyanskii and recovered by Mirlin and Fyodorov\cite{Jacquod97,Mirlin97}.

However, fig. \ref{PS_echelle} shows that the crossover is not uniquely driven by the ratio $U/\Delta_c$, but it also depends 
on the density of  states $\delta$.
For given values of $U$ and $\Delta_c$, when $\Delta_c/\delta$ increases, the transition is faster. This may appear surprising since the density of non-zero coupling elements decreases. However, at the same time, the distance between levels decreases, so that mixing neighboring levels becomes more efficient.

A level of the generation $(2n^*-1)$ is typically coupled to
 a level 
of generation $(2 n^* +1)$ at a distance 
$\Delta_c \gg \delta$. The indirect coupling between two neighboring levels of the same generation (distant of $\delta$)
 typically necessitates
high orders in perturbation $U (U/\Delta_c)^p$ where  $p$ is of order $\sqrt{\Delta_c/\delta}$ . 
The mixing of these neighboring levels and the transition to WD thus involve many such high order 
processes\cite{Silvestrov97}. We have not yet succeeded in finding the correct criterion which involves both energy scales $\Delta_c$ and $\delta$.

    \vspace{0.6cm}

%\newline
{\small 
%%%%%%%%%%%%%%%%%%%%%%%%%%%%%%%%%%%%%%%%%%%%%%%%%%%%%%%%%%%%%%%%%%%%%%%%%% 
%%% acknowledgements, if any; delete otherwise 
We have benefited from useful discussions with 
E. Bogomolny, 
O. Bohigas, P. Jacquod, J.-L. Pichard, B. Shklovskii, 
D. Shepelyansky and P. Walker. M.P. acknowledges financial support from the D.G.A. G.M. acknowledges the hospitality of ICTP, Trieste. Part of the numerical calculations have been performed using IDRIS facilities.
    }

\end{document}